\documentclass{article}
\usepackage{amsmath}
\usepackage{graphicx}
\usepackage{amsfonts}
\usepackage{amssymb}
\usepackage{geometry}
\usepackage{float}
\usepackage{bm}
\usepackage{color}
\usepackage{cite}
\usepackage{subfigure}

\begin{document}

\title{Thermal Properties of Relativistic Dunkl Oscillators}
\author{B. Hamil\thanks{%
hamilbilel@gmail.com} \\
D\'{e}partement de TC de SNV, Universit\'{e} Hassiba Benbouali, Chlef,
Algeria. \and B. C. L\"{u}tf\"{u}o\u{g}lu\thanks{%
bekircanlutfuoglu@gmail.com} \\
Department of Physics, University of Hradec Kr\'{a}lov\'{e}, \\
Rokitansk\'{e}ho 62, 500 03 Hradec Kr\'{a}lov\'{e}, Czechia.}
\date{}
\maketitle

\begin{abstract}
For a better understanding of the physical systems, even at the quantum
level, the thermal quantities can be investigated. Recently, we realized
that the parity of the system can also be examined simultaneously, by
substituting the Dunkl operator to the ordinary differential operator in
quantum mechanics. In this manuscript, we consider two relativistic
Dunkl-oscillators and investigate their thermal quantities with well-known
statistical methods. Besides, we establish a relationship between the
Dunkl-Dirac oscillator and the Dunkl-Anti-Jayne-Cummings model by defining
the Dunk creation and annihilation operators. Therefore, we conclude that
our model can be regarded as an appropriate scenario for the theory of an
open quantum system coupled to a thermal bath.
\end{abstract}

\section{Introduction}

In the middle of the last century, Wigner wrote a very interesting article
discussing whether the equations of motion of a quantum mechanical system
can determine the commutation relations \cite{Wigner1950}. Wigner sought the
answer to this question on free particle and harmonic oscillator systems.
Impressed by Wigner's question, Yang solved the one-dimensional quantum
harmonic oscillator problem the following year by defining an additional
reflection operator in the usual Heisenberg algebra \cite{Yang1951}. This
deformed algebra can be given in the form of \cite{Chung2021} 
\begin{eqnarray}
[\hat{x},\hat{p}]=i (1+\mu \hat{R}),
\end{eqnarray}
where $\mu$ is the Wigner parameter, $\hat{R}$ is the reflection operator
and $\hbar=1$. In this deformed algebra the following relations 
\begin{eqnarray}
\hat{R}\hat{x}=-\hat{x} \hat{R} \quad,\quad \hat{R}\hat{p}=-\hat{p} \hat{R}
\quad,\quad \hat{R}\hat{R}=1 \quad, \quad \hat{R}f\left( x\right) = f\left(
-x\right) .
\end{eqnarray}
are satisfied. It is worth noting that the position representation of this
deformed algebra is not unique. An interesting position representation is
written in terms of the Dunkl operator \cite{Dunkl1989} as 
\begin{eqnarray}
\hat{p}\equiv \frac{1}{i} \hat{D}=\frac{1}{i}\bigg[\frac{d}{dx}+\frac{\mu }{x%
}\left( 1-\hat{R}\right)\bigg],  \label{dunkl}
\end{eqnarray}
where, $\hat{p}$ denotes the usual quantum mechanical momentum operator. A
Dunkl operator, $\hat{D}$, can be defined as a linear superposition of
differential and difference operators \cite{Granados1} for use in various
problems in mathematics \cite{Heckman, DunklXu, Bie} and theoretical physics 
\cite{CMS1, CMS2, Genest1, Genest2, Genest3, Genest4}. Especially in recent
years, we observe that this interest has increased in the context of
relativistic and non-relativistic quantum mechanical problems \cite%
{Sargolzaeipor2018, Ghaz2019, Mota1, CHU1, CHU2020, KIM2020, sta1, sta2 ,
Mota2, Mota3, Merad, Hamil2022}.

In physics, there are various well-known oscillator models, namely
Klein-Gordon, Dirac, Duffin-Kemmer-Petiau, Pauli, or harmonic oscillators.
They are often treated as toy models in research to describe nature. All
these oscillator systems basically differ from each other according to
whether the model is relativistic or not and the spins of the particles.
However, none of them takes the parity of the particles into account. Since
the Dunkl operator carries a parity operator, it can allow obtaining
additional information in the classical solutions. According to this
motivation, some authors already examined Dunkl-Dirac \cite%
{Sargolzaeipor2018,Mota1}, Dunkl-Klein-Gordon \cite{Mota2, Hamil2022},
Dunkl-pseudo harmonic \cite{Mota3} and Dunkl-Duffin-Kemmer-Petiau \cite%
{Merad} oscillators under various scenario and dimensions \cite{Mohadesi}.

On the other hand, especially in the last two decades, we see an increase in
the number of studies discussing the thermal properties of systems in
addition to their quantum structure. For example, thermal properties of one,
three, and fractional Dirac oscillators are investigated in \cite{Pac1,
Pac2, Tdo}. Similarly, thermodynamics quantities of Klein-Gordon,
Duffin-Kemmer-Petiau, harmonic, and Pauli oscillators are derived in \cite%
{Boumali2015, Arda , Samira, Dong, Heddar},
respectively. Last year, Dong et al. studied the Dunkl-Schr\"odinger
equation with a generalized form of Dunkl operator under the effect of the
harmonic oscillator potential energy \cite{sta2}. There, they showed how the
thermal quantities of positive and negative parity systems by comparing the
thermodynamic functions. To our best knowledge, any similar investigation
has not been executed in any Dunkl-relativistic systems. With this
motivation in this manuscript, we consider one-dimensional
Dunkl-Klein-Gordon and Dunkl-Dirac oscillators and intend to obtain their
thermodynamic functions.

We prepare the manuscript as follows: In the next section we derive the
solutions of Dunkl-Klein-Gordon and Dunkl-Dirac oscillators. In section 3,
we give the methodology of cons traction of the partition and thus the
thermal quantity functions. As a numerical example, we study the Dunkl-Dirac
oscillator. After we obtain the Helmholtz free energy, entropy, internal
energy and heat capacity functions, we depict them for the comparison of
positive and negative parity systems. We conclude the manuscript with a
brief section.

\section{One-dimensional relativistic Dunkl-oscillators}

In this section, we briefly introduce two relativistic oscillators, namely
Dunkl-Klein-Gordon and Dunkl-Dirac oscillators, and discuss their solutions
in one spatial dimension.

\subsection{Dunkl-Klein-Gordon oscillator}

\begin{figure*}[htb]
\resizebox{\linewidth}{!}{\includegraphics{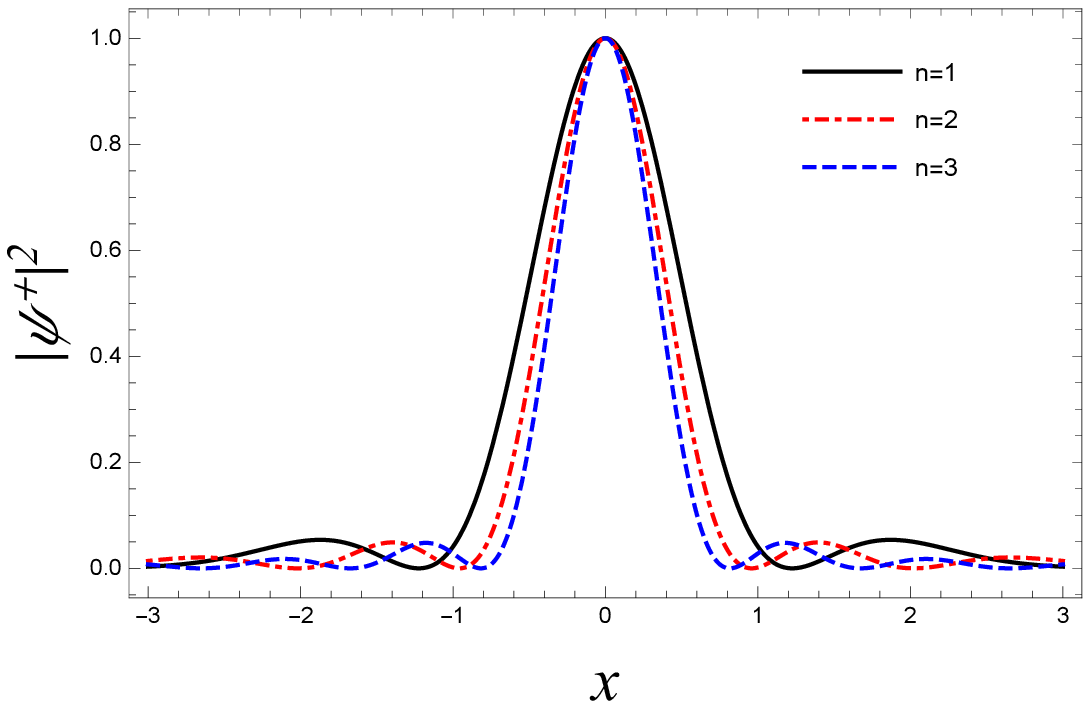},%
\includegraphics{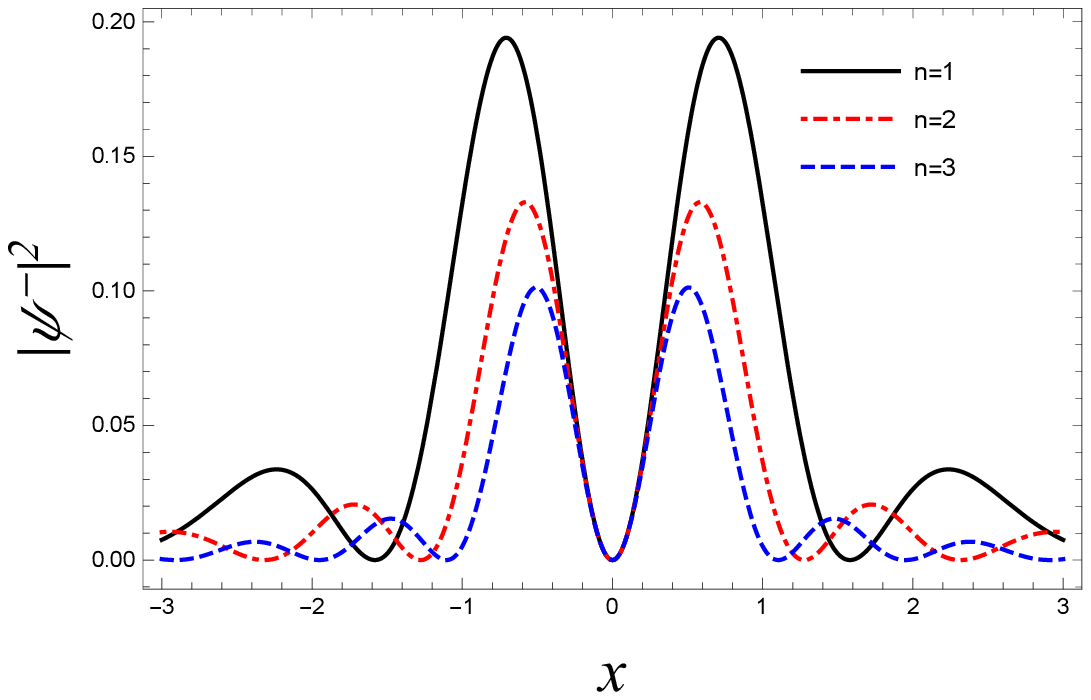}}
\caption{Reduced probability densities versus coordinate.}
\label{Fig1}
\end{figure*}

One-dimensional stationary Dunkl-Klein-Gordon oscillator is given in the
form of 
\begin{equation}
\left[ E^{2}-\left( \frac{1}{i}\hat{D}+im\omega \hat{x}\right) \left( \frac{1%
}{i}\hat{D}-im\omega \hat{x}\right) -m^{2}\right] \psi \left( x\right) =0,
\label{KG}
\end{equation}%
where $m$ and $\omega$ denote the mass and frequency of the oscillator,
respectively. Substituting the Dunkl derivative given in Eq. \eqref{dunkl}
in Eq. \eqref{KG}, we get 
\begin{equation}
\hat{\mathcal{H}}\psi \left( x\right) =0,  \label{H}
\end{equation}%
with the Dunkl-Klein-Gordon Hamilton operator 
\begin{equation}
\hat{\mathcal{H}}=\frac{d^{2}}{dx^{2}}+\frac{2\mu }{x}\frac{d}{dx}-\frac{\mu 
}{x^{2}}\left( 1-\hat{R}\right) -m^{2}\omega ^{2}x^{2}+m\omega \left( 1+2\mu 
\hat{R}\right) +E^{2}-m^{2}.
\end{equation}%
Since the Dunkl-Hamiltonian commutes with the reflection operator, they
should have a common eigenbasis. Therefore, they can be diagonalized
simultaneously. So, the eigenfunction, $\psi \left( x\right) $, can be
selected to have a definite parity ${\hat{R}}\psi \left( x\right)
=s\psi \left( x\right) $ with $s=\pm $. As a result, Eq. (\ref{H}) appears as%
\begin{equation}
\left[ \frac{d^{2}}{dx^{2}}+\frac{2\mu }{x}\frac{d}{dx}-\frac{\mu }{x^{2}}%
\left( 1-s\right) -m^{2}\omega ^{2}x^{2}+m\omega \left( 1+2\mu s\right)
+E^{2}-m^{2}\right] \psi ^{s}\left( x\right) =0.  \label{DKG1}
\end{equation}
The solution of Eq. \eqref{DKG1} is given in many references \cite%
{Mota3,Genest1,Genest2,CHU1}, therefore we address it here briefly. At
first, we introduce a new variable by $y=m\omega x^{2}$, and then, we follow
the ansatz: $\psi ^{s}=y^{\frac{1-s}{4}}e^{-\frac{y}{2}}\Psi ^{s}\left(
y\right) $. After straightforward calculation, we find 
\begin{equation}
\left[ y\frac{d^{2}}{dy^{2}}+\left( 1-\frac{s}{2}+\mu \right) \frac{d}{dy}-%
\frac{y}{4}+\frac{\left[ 1+2\mu s\right] }{4}+\frac{E^{2}-m^{2}}{4m\omega }%
\right] \Psi ^{s}\left( y\right) =0.
\end{equation}%
This differential equation is identified as the confluent hypergeometric
equation, whose solution can be expressed in terms of the confluent
hypergeometric functions, $\mathbf{F}\left( a,b;y\right) $, \cite{grad} 
\begin{equation}
\Psi _{n}^{s}\left( y\right) =\mathcal{C}\mathbf{F}\left( \frac{\left( 2\mu
+1\right) \left( 1-s\right) }{4}-\frac{E^{2}-m^{2}}{4m\omega },1-\frac{s}{2}%
+\mu ;y\right) .
\end{equation}
Therefore, the wave function reads 
\begin{eqnarray}
\Psi_n^s(x)&=& \mathcal{C} (m\omega x)^{\frac{1-s}{2}} e^{-\frac{m \omega x^2%
}{2}}\mathbf{F}\left( \frac{\left( 2\mu +1\right) \left( 1-s\right) }{4}-%
\frac{E^{2}-m^{2}}{4m\omega },1-\frac{s}{2}+\mu ;m \omega x^2 \right).
\end{eqnarray}
The confluent hypergeometric function reduces to a polynomial of degree $n$
in $y$, if the first argument is equal to a negative integer. 
\begin{equation}
\frac{\left( 2\mu +1\right) \left( 1-s\right) }{4}-\frac{E^{2}-m^{2}}{%
4m\omega }=-n.
\end{equation}%
This fact gives the energy quantization 
\begin{eqnarray}
\frac{E_{n}^s}{m} &=&\pm \sqrt{4nr +2r\left( \mu +\frac{1}{2}\right) \Big( %
1-s\Big) +1}{\color{red},}  \label{spec}
\end{eqnarray}
where $r\equiv \omega/m$. We see that the energy spectrum, which depends on
the Wigner parameter, differs according to the parity values. We note in the
absence of $\mu$ and $s$, we obtain the standard result given in \cite{Bruce}%
. In Fig. \ref{Fig1} we plot the first three excited states' probability
densities versus coordinate for positive and negative parities.

\subsection{Dunkl-Dirac oscillator}

\begin{figure*}[htb]
\centering{} \resizebox{0.5\linewidth}{!}{\includegraphics{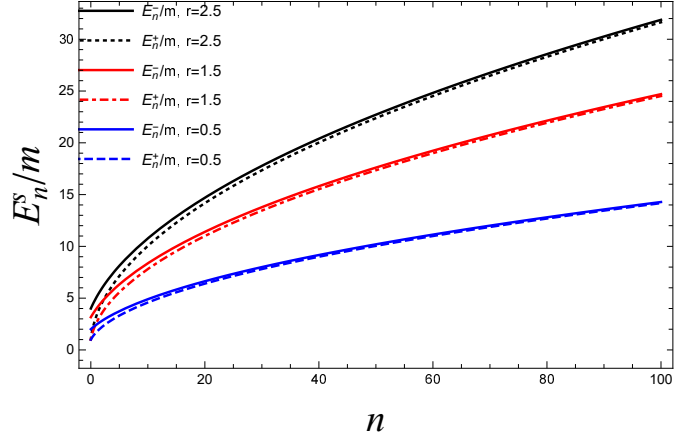}}
\caption{Dunkl-relativistic energy spectra versus the node numbers, where $r=%
\protect \omega/m$.}
\label{Fig2}
\end{figure*}

Next, we consider one-dimensional stationary Dunkl-Dirac oscillator, which
is given in the form of

\begin{eqnarray}
\hat{\mathcal{H}}_D\psi=E^s\psi^s,  \label{DD}
\end{eqnarray}
where 
\begin{equation}
\hat{\mathcal{H}}_D=\left[ \alpha _{x}\left( \frac{1}{i}\hat{D}-i\beta
m\omega \hat{x}\right) +\beta m\right],  \label{ham}
\end{equation}
is the Dunkl-Dirac Hamilton operator, $\psi^s$ is a two-component spinor, $%
\psi =\left( 
\begin{array}{c}
\Phi \\ 
\digamma%
\end{array}%
\right), $ and $\alpha _{x}$ and $\beta $ are the Dirac matrices. Here, we
take the following form 
\begin{equation}
\alpha _{x}=\left( 
\begin{array}{cc}
0 & 1 \\ 
1 & 0%
\end{array}%
\right), \quad \quad \beta =\left( 
\begin{array}{cc}
1 & 0 \\ 
0 & -1%
\end{array}%
\right),
\end{equation}%
and obtain the following decoupled set of equations out of Eq. (\ref{ham}) 
\begin{eqnarray}
\left( \frac{1}{i}D+im\omega x\right) \digamma &=&\left( E^s-m\right) \Phi ,
\label{e1} \\
\left( \frac{1}{i}D-im\omega x\right) \Phi &=&\left( E^s+m\right) \digamma .
\label{e2}
\end{eqnarray}
After performing a simple algebra between Eqs. (\ref{e1}) and (\ref{e2}), we
find the stationary Dirac equation for the upper spinor component as: 
\begin{equation}
\left[ \left( D-m\omega x\right) \left( D+m\omega x\right) +{E^s}^{2}-m^{2}%
\right] \Phi =0.  \label{Dir}
\end{equation}
Since Eq. \eqref{Dir} is similar to Eq. \eqref{KG}, we immediately can write
the solution for the upper component of the spinor as 
\begin{equation}
\Phi ^{s}=\mathcal{N}_1(m\omega x)^{\frac{1-s}{2}} e^{-\frac{m \omega x^2}{2}%
}\mathbf{F}\left( -n,1-\frac{s}{2}+\mu ;m \omega x^2 \right).  \label{wave}
\end{equation}%
Thus, the spinor solution reads 
\begin{equation}
\psi _{n}^{s} =\mathcal{N}_{s}(m\omega x)^{\frac{1-s}{2}} e^{-\frac{m \omega
x^2}{2}}\left( 
\begin{array}{c}
1 \\ 
\frac{-i}{\left( E+m\right) }\left[ \frac{d}{dx}+\frac{\left( \mu +\frac{1}{2%
}\right) \left( 1-s\right) }{x}\right]%
\end{array}%
\right) \mathbf{F}\left( -n,1-\frac{s}{2}+\mu ;m\omega x^{2}\right) .
\end{equation}
The energy quantization leads to the same energy function given in Eq. %
\eqref{spec}. We depict the energy spectrum function versus the node numbers
in Fig. \ref{Fig2}. We observe that the energy spectrum of even and odd
parity particles differs relatively more in the ground states. It is worth
noting that for $\mu=s=0$, the energy function reduces to the one given in 
\cite{Dominguez}. 

In order to establish a connection between the Dunkl-Dirac oscillator and
quantum optics, we introduce the Dunkl creation $\hat{a}^\dag_D$ and
annihilation $\hat{a}_D$ operators 
\begin{eqnarray}
\hat{a}^\dag_D&=&\frac{1}{\sqrt{2m\hbar \omega}}\left(m\omega \hat{x}-\hbar 
\hat{D}\right),  \label{cre} \\
\hat{a}_D&=&\frac{1}{\sqrt{2m\hbar \omega}}\left(m\omega \hat{x}+\hbar \hat{D%
}\right),  \label{ann}
\end{eqnarray}
which satisfy 
\begin{eqnarray}
\left[\hat{a}^\dag_D,\hat{a}_D\right]&=&1+2\mu \hat{R}.
\end{eqnarray}
By substituting Eqs. \eqref{cre} and \eqref{ann} into Eq. \eqref{DD}, we
express the Dunkl-Dirac Hamiltonian in the form of 
\begin{eqnarray}
\hat{\mathcal{H}}_D&=&g \left(\sigma^- \hat{a}_D + \sigma^+ \hat{a}%
^\dag_D\right)+ m\sigma_z,
\end{eqnarray}
where $\sigma^\mp=\frac{1}{2}\left(\sigma_x\mp \sigma_y\right)$. It is worth
noting that the mapped Hamiltonian corresponds to the
Dunkl-Anti-Jaynes-Cummings (DAJC) model and it reduces to the ordinary
Anti-Jaynes-Cummings (AJC) model when the Wigner parameter is taken as zero.

\section{Thermodynamics of relativistic Dunkl-oscillators}
In this section, we examine the thermal properties of the relativistic
Dunkl-oscillators. Since the energy spectrum function of Klein-Gordon and Dirac oscillators are the same, their thermal properties are also similar \cite{Boumali2015}. However, in the Dunkl case, the relativistic energy spectra of odd and even parity particles differ from each others, therefore, their partition
functions, thus, thermal properties differ.

\begin{figure*}[htb]
\resizebox{\linewidth}{!}{\includegraphics{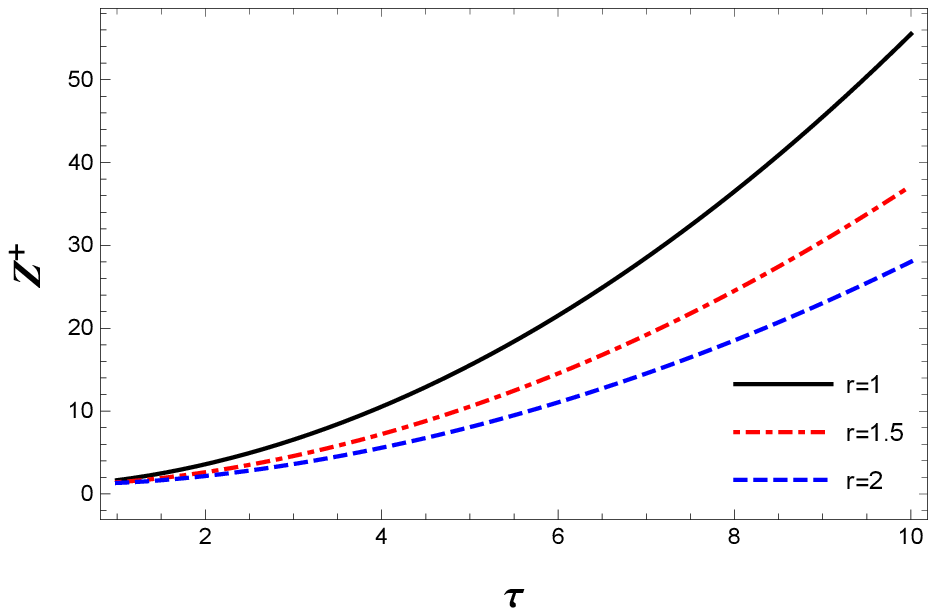},\includegraphics{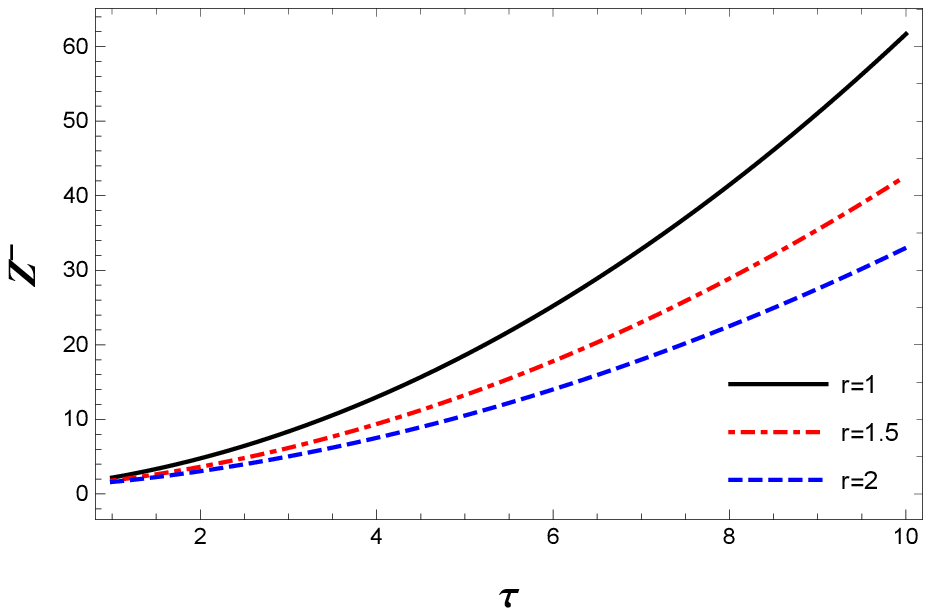}}
\caption{Dunkl-partition functions versus reduced temperature.}
\label{F3}
\end{figure*}

We start by assuming the system is in thermal equilibrium with a thermal bath at the
temperature $T$. Then, we write the canonical ensemble partition function as 
\begin{equation}
Z^{s}=\sum \limits_{n=0}e^{-\frac{E_{n}^{s}-E_{0}^{s}}{K_{B}T}},  \label{Ser}
\end{equation}%
where $K_{B}$ is the Boltzmann constant, and $E_{0}^{s}$
is the ground-state energy corresponding to $n=0.$ Here, we have to restrict ourselves with the stationary state’s positive energy solutions, since the partition function does not converge with the negative energy solutions in canonical ensemble. As our initial evaluation, let us test the convergence of the series given in Eq. (\ref%
{Ser}) by using the integral formula%
\begin{equation}
\int_{0}^{+\infty }e^{-\frac{1}{\tau }\sqrt{ax+b}}dx=\frac{2\tau ^{2}}{a}%
\left( 1+\tau \sqrt{b}\right) e^{-\frac{1}{\tau }\sqrt{b}},\text{ \  \  \  \  \
for }\tau >0.
\end{equation}%
This implies that the partition function is convergent. In order
to evaluate this function, we use the Euler-MacLaurin formula defined with \cite{Pacheco}
\begin{equation}
\sum \limits_{n=0}f\left( n\right) =f\left( 0\right) +\int_{0}^{+\infty
}f\left( x\right) dx-\sum \limits_{p=1}\frac{B_{2p}}{\left( 2p\right) !}%
f^{\left( 2p-1\right) }\left( 0\right) ,  \label{ABC}
\end{equation}%
where $B_{2p}$ are the Bernoulli numbers, i.e. $B_2=\frac{1}{6}$, $B_4=-\frac{1}{30}$, and $B_6=\frac{1}{42}$ . Therefore, Eq. \eqref{ABC} reads
\begin{equation}
\sum \limits_{n=0}f\left( n\right) =f\left( 0\right) +\int_{0}^{+\infty
}f\left( x\right) dx-\frac{1}{12}f^{\left( 1\right) }\left( 0\right) +\frac{1%
}{720}f^{\left( 3\right) }\left( 0\right) -\frac{1}{30240}f^{\left( 5\right)
}\left( 0\right) +....
\end{equation}%
Here, $f^{\left( n\right) }$ is the derivative of order $n$. In our case, the determination of the partition function can only be carried out on numerical methods. We find the partition function in the form of
\begin{equation}
Z^{s}=\frac{1}{2}+\frac{\sqrt{\alpha_s }\tau }{2r}+\frac{\tau ^{2}}{2r}+\frac{r}{6%
\sqrt{\alpha_s }\tau }+\mathcal{O}\left( \frac{1}{\tau ^{3}}\right) .\label{ZS}
\end{equation}
Here, we use the following dimensionless parameters,%
\begin{equation}
\alpha_s \equiv 2r\left( \frac{1}{2}+\mu \right) \left( 1-s\right) +1,\quad\quad
\tau \equiv \frac{K_{B}T}{m}=\frac{T}{T_0}, \quad\quad 
r \equiv \frac{\omega }{m},
\end{equation}
where $T_0=\frac{m}{K_B}$ is the characteristic temperature that splits the range of temperature to very low temperature, $T<<T_0$, and very high temperature, $T>>T_0$, regions \cite{Boumali2013}. We observe that only in the odd case the Wigner parameter value takes a role. Also, it's worth mentioning that in the high temperature regime, $T\rightarrow
\infty $, the Dunkl parameter's contribution remains negligible, and the
partition functions become unique and indistinguishable. \begin{equation}
Z^{+}=Z^{-}=\frac{\tau ^{2}}{2r}.
\end{equation}
Before we derive the thermal quantities, we depict the Dunkl-partition functions versus reduced temperature in Fig. \ref{F3}.  It is worth noting that, in the low temperature region, $(T<T_0)$, the higher order terms neglected in Eq. \eqref{ZS} will come to dominate. Therefore, we have to show the characteristic behaviour of the thermal quantities in the interval of $\tau \geq 1$.  Hereafter, we take $K_{B} = m = 1 $ in the rest of the manuscript, and we examine three different oscillator frequencies, $r=1$, $r=1.5$, $r=2$. Furthermore,
in the odd partition function, we assume $\mu =1/2$.
We observe that the partition function of even and odd parity systems
increases monotonically with temperature and for a fixed temperature the
partition function decreases when the oscillator frequency grows. Moreover,
we notice that the partition function of even and odd parity systems differs
at relatively small temperatures. 

\begin{figure*}[htb]
\resizebox{\linewidth}{!}{\includegraphics{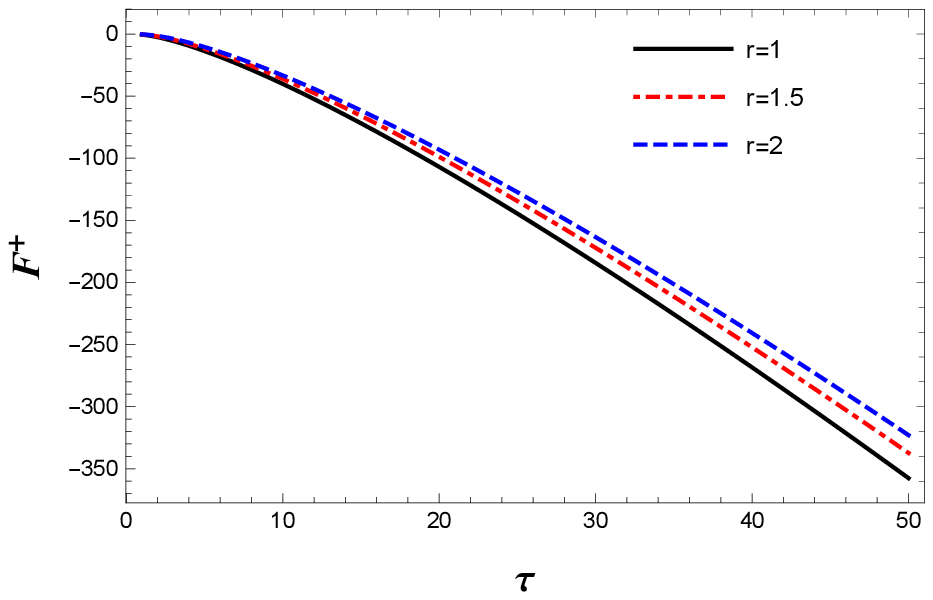},\includegraphics{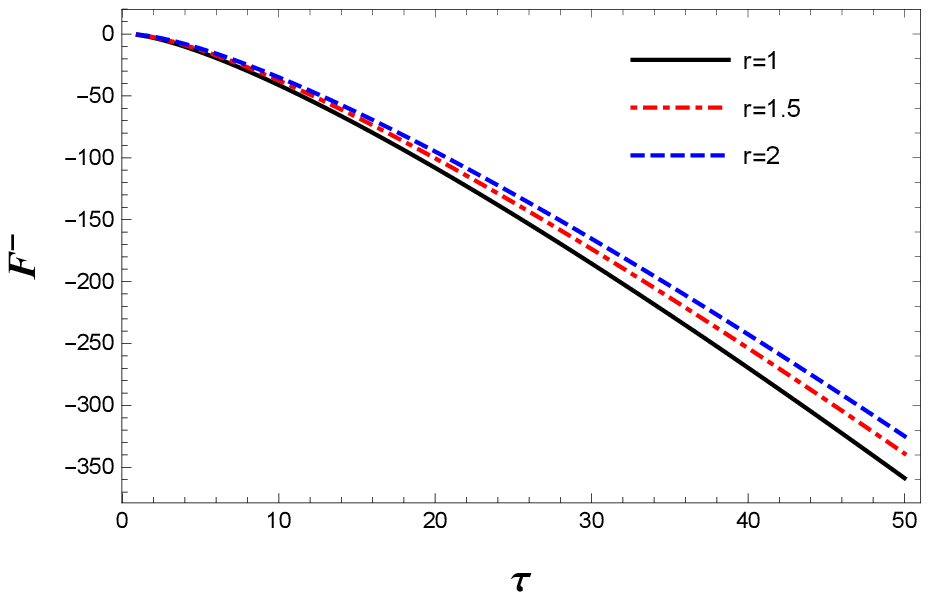}}
\caption{Dunkl-Helmholtz free energy versus reduced temperature.}
\label{F4}
\end{figure*}

After determining the partition function, we derive the Helmholtz free energy, entropy, mean energy and specific heat
functions of the relativistic Dunkl-oscillators. At first, we use 
\begin{equation*}
F^{s}=-\tau \ln Z^{s},
\end{equation*}%
\newpage
to derive the Dunkl-Helmholtz free energy. We find%
\begin{equation}
F^{s}=-\tau \ln \left[ \frac{1}{2}+\frac{1}{2r}\left( \tau ^{2}+\tau \sqrt{%
\alpha_s }\right) +\frac{r}{6\tau \sqrt{\alpha_s }}\right] .
\end{equation}%
In Fig. \ref{F4},  we demonstrate the plots of the Dunkl-Helmholtz free energy functions of the relativistic oscillators for the considered oscillator frequencies. In
both systems, when the oscillator has greater frequency, then system's
thermal function gets greater values.

\begin{figure*}[htb]
\resizebox{\linewidth}{!}{\includegraphics{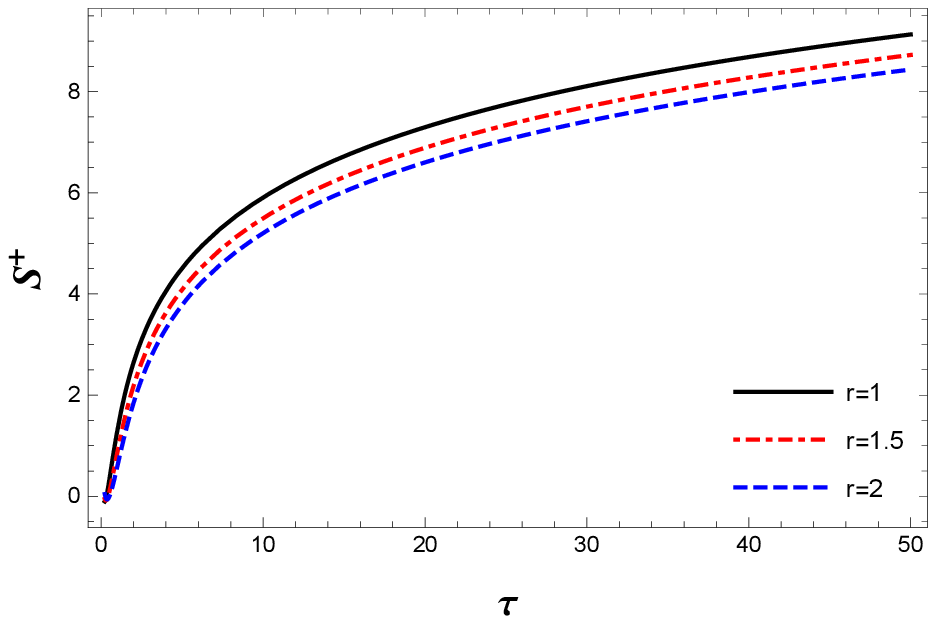},\includegraphics{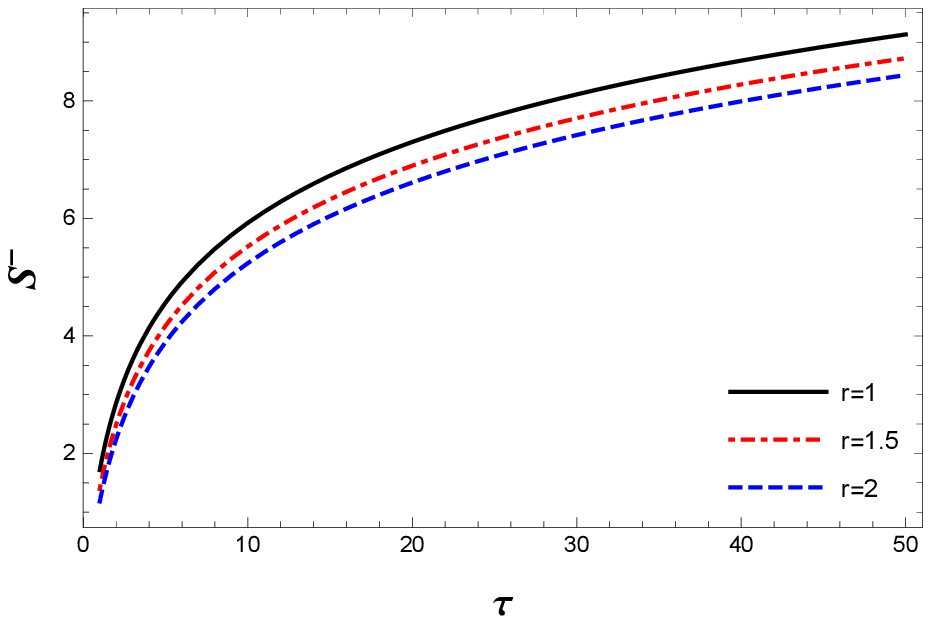}}
\caption{Dunkl-entropy versus reduced temperature.}
\label{F5}
\end{figure*}

Next, we derive the reduced Dunkl-entropy function via 
\begin{equation*}
S^{s}=\ln Z^{s}+\tau \frac{\partial }{\partial \tau }\ln Z^{s}.
\end{equation*}%
For even and odd parity systems, we obtain
\begin{equation}
S^{s}=\ln \left[ \frac{1}{2}+\frac{1}{2r}\left( \tau ^{2}+\tau \sqrt{\alpha_s }%
\right) +\frac{r}{6\tau \sqrt{\alpha_s }}\right] +\tau \left[\frac{%
\sqrt{\alpha_s }+2\tau }{2r}-\frac{r}{6\tau ^{2}\sqrt{\alpha_s }}\right]\left[\frac{1}{2}+%
\frac{r}{6\tau \sqrt{\alpha_s }}+\frac{\tau \sqrt{\alpha_s }+\tau ^{2}}{2r}\right]^{-1}.
\end{equation}%
Then, we demonstrate the Dunkl-entropy functions in Fig. \ref{F5}. We
observe that in both systems the Dunkl-entropy function saturates at lower values for
greater oscillator frequencies.

\begin{figure*}[htb]
\resizebox{\linewidth}{!}{\includegraphics{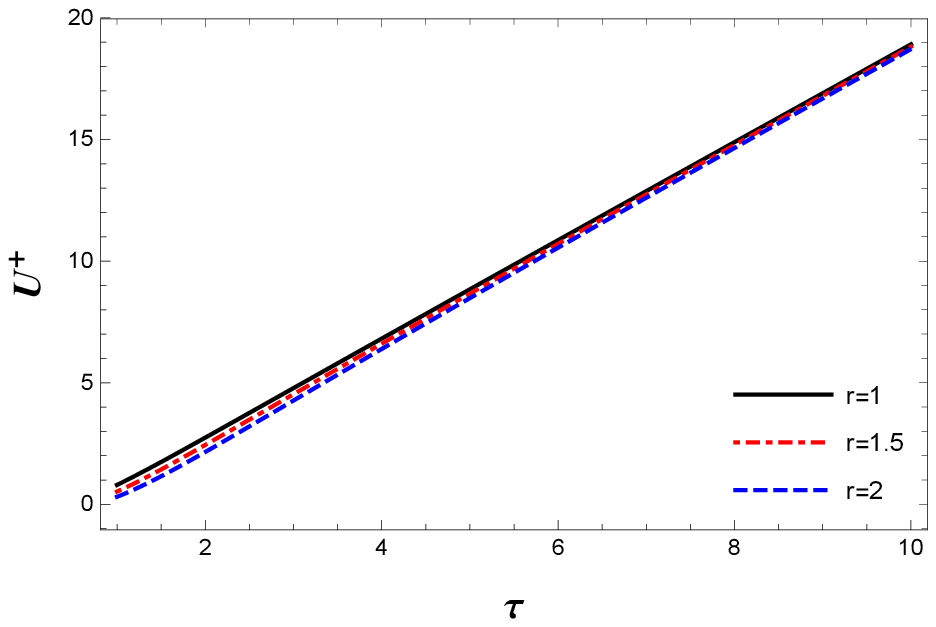},\includegraphics{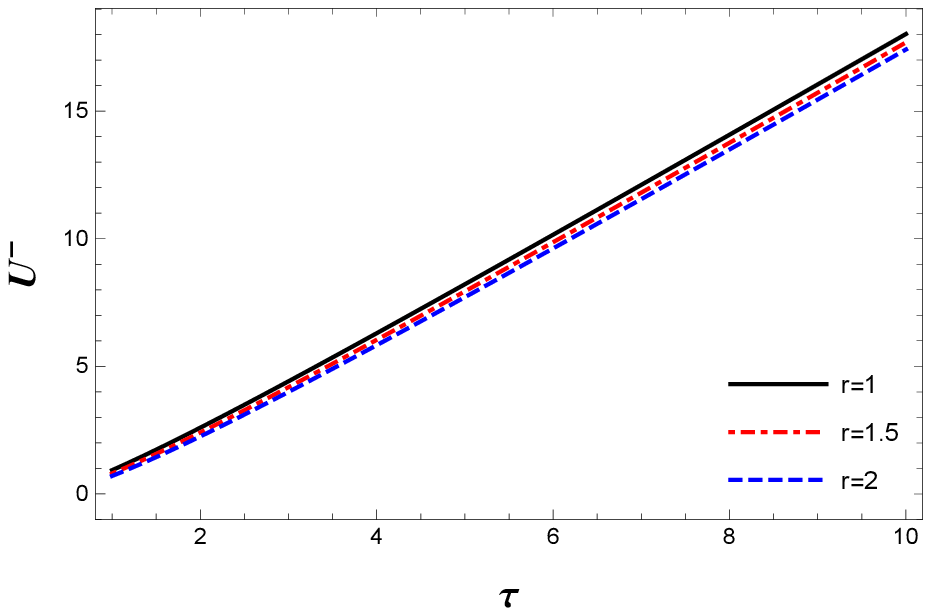}}
\caption{Dunkl-mean energy versus reduced temperature.}
\label{F6}
\end{figure*}

Next, we examine the mean energy function by using 
\begin{equation*}
U^{s}=\tau ^{2}\frac{\partial }{\partial \tau }\ln Z^{s}.
\end{equation*}%
We find 
\begin{equation}
U^{s}=\tau^{2}\left[\frac{\sqrt{\alpha_s }+2\tau }{2r}-\frac{r}{6\sqrt{\alpha_s 
}\tau ^{2}}\right]
\left[\frac{1}{2}+\frac{r}{6\sqrt{\alpha_s }\tau }+\frac{\sqrt{\alpha_s }%
\tau +\tau ^{2}}{2r}\right]^{-1}.
\end{equation}%
In Fig. \ref{F6}, we present the characteristic behavior of the Dunkl-mean energy functions. We observe that at high temperatures Dunkl-mean energy functions tend to the same value, $U^s \rightarrow 2 T$, in the even and odd relativistic cases.

\begin{figure*}[htb]
\resizebox{\linewidth}{!}{\includegraphics{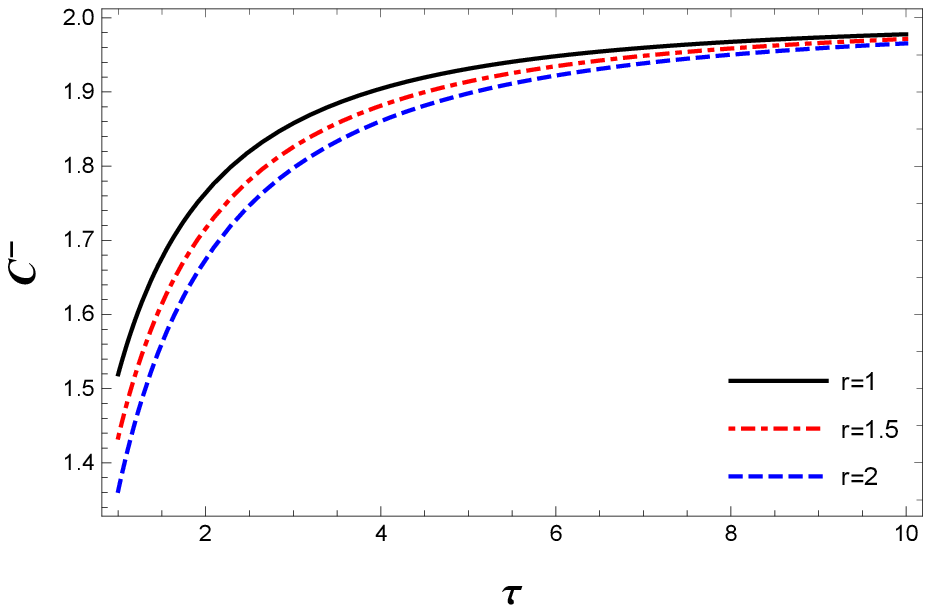},\includegraphics{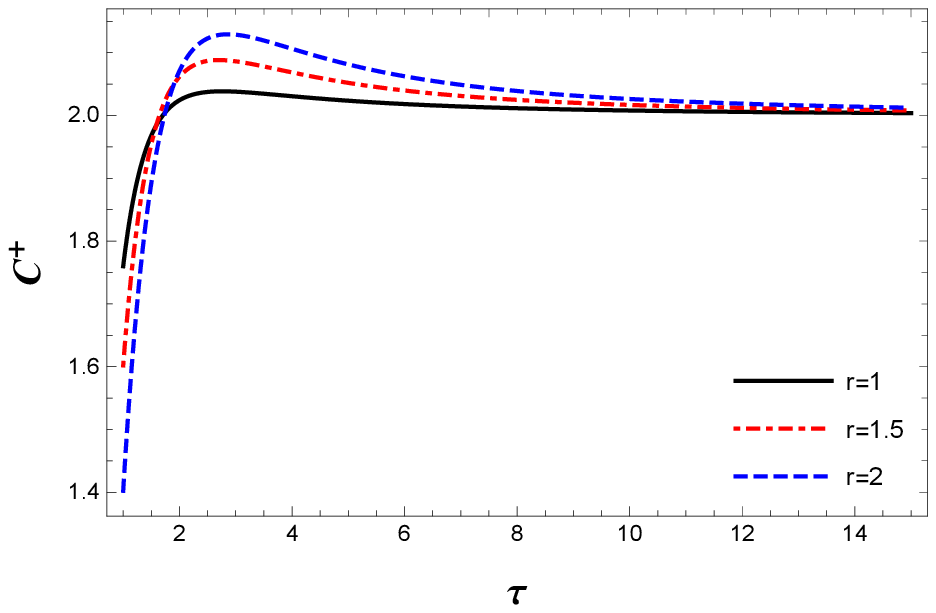}}
\caption{Dunkl-heat capacity versus reduced temperature.}
\label{F7}
\end{figure*}

Finally, we derive the reduced Dunkl-heat capacity function by using 
\begin{equation*}
C^{s}=2\tau \frac{\partial }{\partial \tau }\ln Z^{s}+\tau ^{2}\frac{%
\partial ^{2}}{\partial \tau ^{2}}\ln Z^{s}.
\end{equation*}%
We obtain 
\begin{equation}
C^{s}=2\tau \frac{\frac{\sqrt{\alpha_s }+2\tau }{2r}-\frac{r}{6\sqrt{\alpha_s }%
\tau ^{2}}}{\frac{\sqrt{\alpha_s }\tau +\tau ^{2}}{2r}+\frac{r}{6\sqrt{\alpha_s }%
\tau }+\frac{1}{2}}+\tau ^{2}\frac{\frac{r}{3\sqrt{\alpha_s }\tau ^{3}}+\frac{1%
}{r}}{\frac{\sqrt{\alpha_s }\tau +\tau ^{2}}{2r}+\frac{r}{6\sqrt{\alpha_s }\tau }%
+\frac{1}{2}}-\tau ^{2}\frac{\left( \frac{\sqrt{\alpha_s }+2\tau }{2r}-\frac{r%
}{6\sqrt{\alpha_s }\tau ^{2}}\right) ^{2}}{\left( \frac{\sqrt{\alpha_s }\tau
+\tau ^{2}}{2r}+\frac{r}{6\sqrt{\alpha_s }\tau }+\frac{1}{2}\right) ^{2}}.
\end{equation}%
Then, we depict the Dunkl-heat functions versus reduced temperature in Fig.%
\ref{F7}.
We see that all relativistic Dunkl-heat capacity functions converge to 2 at high temperatures. On the other hand, at low temperatures, the Dunkl-heat capacity functions behave differently for the even and odd relativistic cases.

\section{Conclusion}

In this manuscript, we examine the thermal properties of relativistic
Dunkl-oscillators. To this end, at first we review the solutions of the
Dunkl-Klein-Gordon and Dunkl-Dirac oscillators in one dimension. Then, we
take the odd and even parity systems in thermal equilibrium and express the
canonical partition functions. After that, we derive the reduced
Dunkl-Helmholtz free energy, Dunk-entropy, Dunkl-mean energy and Dunkl-heat
capacity functions. In even and odd parity systems, we demonstrate all these
thermal quantities versus the reduced temperature.

Besides all, we mapped the Dunkl-Dirac oscillator Hamiltonian onto
Dunkl-Anti-Jaynes-Cummings model of quantum optics. Quantum optics theory
offers one of the first test fields of open quantum systems. Basically, an
open quantum system deals with the problems of dephasing and damping in the
considered system with the claim that all real systems are open and interact
with the surroundings \cite{Samira, Dong}. Therefore, the Dunkl-Dirac
oscillator can be regarded as an appropriate scenario for the theory of an
open quantum system coupled to a thermal bath.

\section*{Acknowledgments}

The authors would like to thank the anonymous referees for their precious
critics and comments. One of the authors of this manuscript, BCL, is
supported by the Internal Project, [2022/2218], of Excellent Research of the
Faculty of Science of Hradec Kr\'alov\'e University.

\section*{Data Availability Statements}

The authors declare that the data supporting the findings of this study are
available within the article.

\end{document}